# Thickness-dependent catalytic activity of hydrogen evolution based on single atomic catalyst of Pt above MXene

Zheng Shu[1] and Yongqing Cai[1,*]

[1]Joint Key Laboratory of the Ministry of Education, Institute of Applied Physics and

Materials Engineering, University of Macau, Taipa, Macau, China

*Corresponding author: yongqingcai@um.edu.mo

## Abstract

Hydrogen as the cleanest energy carrier is a promising alternative renewable resource to fossil fuels. There is an ever-increasing interest in exploring efficient and cost-effective approaches of hydrogen production. Recent experiments have shown that single platinum atom immobilized on the metal vacancies of MXenes allows a high-efficient hydrogen evolution reaction (HER). Here using *ab initio* calculations, we design a series of substitutional Pt-doped $Ti_{n+1}C_nT_x$ ($Ti_{n+1}C_nT_x$-$Pt_{SA}$) with different thicknesses and terminations ($n$ = 1, 2 and 3, $T_x$ = O, F and OH), and investigate the quantum-confinement effect on the HER catalytic performance. Surprisingly, we reveal a strong thickness effect of the MXene layer on the HER performance. Amongst the various surface-terminated derivatives, $Ti_2CF_2$-$Pt_{SA}$ and $Ti_2CH_2O_2$-$Pt_{SA}$ are found to be the best HER catalysts with the change of Gibbs free energy $\Delta G_{*H} \sim 0$ eV, complying with the thermoneutral condition. The *ab initio* molecular dynamics simulations reveal that $Ti_2CF_2$-$Pt_{SA}$ and $Ti_2CH_2O_2$-$Pt_{SA}$ possess a good thermodynamic stability. The present work shows that the HER catalytic activity of the MXene is not solely governed by the local environment of the surface such as Pt single atom. We point out the critical role of thickness control and surface decoration of substrate in achieving a high-performance HER catalytical activity.

## 1. Introduction

As a carbon-free fuel, hydrogen is a clean energy source and a promising alternative to fossil fuel [1-3]. The hydrogen production from water splitting via electrocatalysts allows the utilization of solar-driven electricity for an indirect solar-chemical energy conversion with zero carbon emission [4, 5]. The cathodic hydrogen reduction reaction (HER) as one of the half-reactions in electrochemical water splitting is a critical process governing the hydrogen production. Until now, the high-cost catalyst like bulk platinum is the most efficient catalysts for HER [6]. The search for high-performance and low-cost electrocatalysts is vital to the scalable application of HER [5, 7]. Developing high-active and cost-effective electrocatalysts with a promoted HER activity has roused the great interest.

With the high surface volume ratio, two-dimensional (2D) materials came into the spotlight as a component of various co-catalysts for HER [8-10]. Since the discovery of graphene, other layered materials analogue of graphene have been successively fabricated and investigated [11-16] with respect to their great potential in mechanical, optoelectronic, electrocatalytic, piezoelectric and thermoelectric applications [9-16]. Shaikh *et al* [17] experimentally synthesized various non-stoichiometric 2D $MoS_2$ nanosheets by hydrothermal route and revealed that sulphur enriched nanosheets show the lower overpotential than equivalent stoichiometric $MoS_2$. Chen *et al* [18] explored the structure-property-activity relationship based on 2D COFs for a photocatalytic HER, and found out that the photocatalytic and electronic properties of 2D COFs can be adjusted by incorporating transition metals into the porphyrin rings. Liang *et al* [19] designed a monolayer $Ni-MoS_2$ to enhance the HER performance over the pristine $MoS_2$ by DFT calculations. Cai *et al* [20] revealed that atomic vacancies and edges in phosphorene play a dominant role in activating the HER. Yu *et al* [21] proved that the HER activity of a metallic 2D $TaS_2$ can be tuned based on different substrates.

MXenes, as the graphene-like 2D transition metal carbides, nitrides or carbonitrides, have a superior metallicity with a chemical composition of $M_{n+1}X_nT_x$ ($n$ can be 1-4, and $T_x$ is the surface termination) [22, 23]. This good metallic conductivity allows a fast flow of conducting carriers (~24000 S cm$^{-1}$ for $Ti_3C_2T_x$), thus rendering them as efficient electrochemical catalysts[24]. Moreover, the properties of MXenes can be significantly modulated by altering surface termination or adopting ordered double transition metals and alloying [25-27], which in turn affects the electrocatalytic performance. Furthermore, the lamellar structure endows MXenes and MXene-based heterostructures an outstanding activity in a variety of electrocatalytic processes, such as HER, oxygen evolution/reduction, nitrogen reduction, carbon dioxide reduction and nitrate reduction to ammonia [28-37]. For instance, Pt-$V_O$-$Ti_2CO_2$ can act as an efficient bifunctional catalyst for oxygen evolution/reduction by tensile strain [38]. Moreover, the HER performance can be strongly enhanced by platinum substitutional doping of cation in the outer layer of MXenes. Zhang *et al* [39] reported that the as-formed Mo vacancies can immobilize Pt atoms for enhancing the MXene's HER activity and reducing the cost of the noble Pt in the catalysts. Jing *et al* [40] figured out that the O vacancies on the surface of $Mo_2TiC_2$ act as the trapping centers of Pt atoms. However, the mechanism of the promoted HER activity of the Pt-doped MXenes is still unclear. The mystery remains about the compositional variance of the performance [41] and the effect of the thickness of the MXene substrate for hosting the single Pt atom.

In this work, 12 kinds of substitutional Pt-doped $Ti_{n+1}C_nT_x$ ($Ti_{n+1}C_nT_x$-$Pt_{SA}$, $n = 1$, 2 and 3, $T_x$ = O, F and OH) are systematically examined toward the potential HER electrocatalysts by using first-principles calculations. We find that the factor of the thickness of $Ti_{n+1}C_nT_x$, which has normally been overlooked before, plays a key role in affecting the HER performance. Through a delicate screening, Pt decorated $Ti_2CF_2$ and $Ti_2CH_2O_2$ are predicted to be the promising HER electrocatalysts with an appropriate free energy of hydrogen adsorption ($\Delta G_{*H}$) of nearly zero. In addition, *ab initio* molecular dynamics (AIMD) was performed to prove that the single Pt atom can be stabilized on Ti vacancy. Our work provides a new perspective to design high-performance HER electrocatalysts for guiding experimental synthesis.

## 2. Computational details

We perform spin-polarized *ab initio* calculations to optimize the structure and obtain total energies by the VASP [42] package. The exchange-correlation interaction between electrons is treated by generalized-gradient approximation with Perdew–Burke–Ernzerhof functional (GGA-PBE) [43]. The interaction between the core and the valence shell is described by the projector augmented wave (PAW) [44] potentials. The van der Waals interaction is also considered by the semiempirical DFT-D3 approach [45]. A cut-off energy of 500 eV in a basis set of plane-wave is adopted for electron wave function extension. The $3 \times 3 \times 1$ supercell of $Ti_{n+1}C_nT_x$ is used for the creation of the defect and the immobilized Pt single atom. In addition, a vacuum layer of ~20 Å is adopted to avoid the interaction between two periodic images along the $Z$ direction. The structures are fully relaxed until the forces acting on each atoms are smaller than 0.02 eV Å$^{-1}$. $15 \times 15 \times 1$ k-points in 2D Brillouin zone sampling are adopted for the relaxation of primitive cell, then the single gamma point is used to optimize $3 \times 3 \times 1$ supercell. $3 \times 3 \times 1$ and $6 \times 6 \times 1$ k-points within the Monkhorst-Pack scheme are used for the calculation of total energies and electronic properties of the supercell. The differential charge density is calculated by $\Delta\rho \ (r) = \rho_{cat+H} \ (r) - \rho_{cat} \ (r) - \rho_H \ (r)$, where $\rho_{cat+H} \ (r)$, $\rho_{cat} \ (r)$ and $\rho_H \ (r)$ are the charge densities of the H adsorbed $Ti_{n+1}C_nT_x$-$Pt_{SA}$, $Ti_{n+1}C_nT_x$-$Pt_{SA}$ and H atom, respectively. Atomic charge transfer is analyzed by Bader charge method [46]. The defect formation energies ($E_f$) can be calculated by the following expression [39]:

$$E_{\text{f}} = E_{\text{defect}} - E_{\text{pure}} - \sum_{\text{i}} n_{\text{i}} \mu_{\text{i}},$$

where $E_{\text{defect}}$ and $E_{\text{pure}}$ are the total energies of defect-contained and pure $\text{Ti}_{n+1}\text{C}_n\text{T}_x$ supercells, respectively. $n_i$ is the number of removed (with a sign of -) or added (with a sign of +) atoms for the $i$ type element, and $\mu_i$ is the chemical potential of $i$ type element of its elemental phase. Besides, the binding energies ($E_{\text{b}}$) of Pt atoms on defect-contained $\text{Ti}_{n+1}\text{C}_n\text{T}_x$ are calculated using the following equation:

$$E_{\text{b}} = E_{\text{defect+Pt}} - E_{\text{Pt}} - E_{\text{defect}},$$

where $E_{\text{defect+Pt}}$, $E_{\text{Pt}}$, $E_{\text{defect}}$ are the total energies of $\text{Ti}_{n+1}\text{C}_n\text{T}_x\text{-Pt}_{\text{SA}}$ monolayer, Pt single atom and $\text{Ti}_{n+1}\text{C}_n\text{T}_x$ with one Ti atomic vacancy. A negative value of $E_{\text{b}}$ indicates a stable adsorption of Pt. The HER activities can be evaluated via the Gibbs free energy change of H adsorption $\Delta G_{\text{H}^*}$ using the standard hydrogen electrode (SHE) model proposed by Nørskov *et al.* [47], which is defined as:

$$\Delta G_{\text{H}^*} = \Delta E_{\text{H}^*} + \Delta E_{\text{ZPE}} - T\Delta S,$$

where $\Delta E_{\text{H}^*}$ can be obtained from the calculations of the total energies of H-adsorbed $\text{Ti}_{n+1}\text{C}_n\text{T}_x\text{-Pt}_{\text{SA}}$, $\text{Ti}_{n+1}\text{C}_n\text{T}_x\text{-Pt}_{\text{SA}}$, and isolated $\text{H}_2$ molecule. $\Delta E_{\text{ZPE}}$ is the correction of zero point energy, and $\Delta S$ is the entropy difference between the gas $\text{H}_2$ and adsorbed H. Both $\Delta E_{\text{ZPE}}$ and $\Delta S$ can be calculated by the vibrational frequency of *H. For example, the vibrational frequency of *H has three eigenvalues: 1429.79, 861.28 and 854.86 cm$^{-1}$ in the case of $\text{Ti}_4\text{C}_3\text{H}_2\text{O}_2\text{-Pt}_{\text{SA}}$. Thus the $E_{\text{ZPE}}$ and $TS$ of *H can be calculated by

$$E_{ZPE} = \frac{1}{2} \sum_i h\, \nu_i,$$

$$-TS = k_B T \sum_i \ln\left(1 - e^{-\frac{h\nu_i}{k_B T}}\right) - \sum_i h\nu_i \left(\frac{1}{e^{\frac{h\nu_i}{k_B T}} - 1}\right).$$

In addition, AIMD simulations are carried out under canonical ensemble (NVT) using the Nosé-Hoover heat bath [48] to validate the thermodynamic stability of the systems.

## 3. Results and discussions

### 3.1. Formation of Ti-vacancy and immobilization of Pt single-atom

A $3 \times 3 \times 1$ supercell of $Ti_{n+1}C_nT_x$ (Figure 1) is constructed to create the Ti vacancy and thereafter host a Pt single atom for catalyzing HER. Both $Ti_2CT_x$ and $Ti_4C_3T_x$ have $P$-3$m$1 space group, whereas $Ti_3C_2T_x$ has $P$-6$m$2 space group. The optimized lattice parameters and layer thickness of this series are presented in Table 1, which agrees well with C2DB database [49]. Noted that the generation of such Ti vacancy in experiments can be realized by etching [39]. Here the generation of the substitutional Pt atom on the surface could occur though anchoring the Pt atom above the Ti vacancy and the process is schematically shown in Figure 2(a). Therefore, to expose the defect sites to immobilize Pt atoms, the formation of a certain amount of Ti vacancy at surface is necessary. The formation of defects not only creates specific sites for Pt immobilization for the HER, but also modulates the charge distribution and electronic properties which also affect the catalytic performance [31, 33]. Here the $E_f$ of Ti vacancy is firstly investigated. The calculated chemical potential $\mu_{Ti}$ at Ti-rich environment is -8.39 eV. As shown in Figure 2(b), the $E_f$ of the out-surface Ti vacancy of $Ti_{n+1}C_nO_2$ is obviously larger than other MXenes considered regardless of the thickness.

Next we calculate the binding energies ($E_b$) of Pt above the Ti vacancy to assess the stability of Ti$_{n+1}$C$_n$T$_x$-Pt$_{SA}$ systems. The surfaces of MXenes are usually terminated with functional groups including O, F, OH, which dramatically change the adsorption, electronic and electrocatalytic properties of MXenes [24, 50, 51]. Table 2 and Figure 2(c) compile the $E_b$ of Pt atom at the Ti vacancy site of the MXene surface which is negative ranging from -9.31 to -4.92 eV, suggesting the presence of the Ti vacancy can stabilize the Pt single atom.

### 3.2. Gibbs free energies of H$^+$ adsorbed on Ti$_{n+1}$C$_n$T$_x$-Pt$_{SA}$

Nørskov *et al* have revealed that $\Delta G_{*H}$ is an effective descriptor for the HER by theoretical and experimental studies [47]. To evaluate the HER activity, we calculated the vibrational frequencies, zero-point energies ($E_{ZPE}$), entropy corrections ($TS$) for adsorbed H on Ti$_{n+1}$C$_n$T$_x$-Pt$_{SA}$, and the corresponding $\Delta G_{*H}$ for these catalytic systems, which are compiled in Table 3. The optimized H adsorption configurations of Ti$_{n+1}$C$_n$T$_x$-Pt$_{SA}$ are presented in Figure 3. As we can see, H tends to preferably adsorb on O atoms of Ti$_{n+1}$C$_n$O$_2$-Pt$_{SA}$ surface while bonding with Pt atom on other Ti$_{n+1}$C$_n$T$_x$-Pt$_{SA}$ (T = F and OH) surfaces. As shown in Figure 4(a), the $\Delta G_{*H}$ on Ti$_{n+1}$C$_n$-Pt$_{SA}$ without terminal groups is up to ~ -0.60 eV, indicative of the suppression of HER activity due to a strong adsorption of H. The value of $\Delta G_{*H}$ varies slightly between -0.64 and -0.57 eV with $n$ changing from 1 to 3. This indicates that for the naked MXene the catalytic activity is only slightly dependent on the thickness and the $Ti_{n+1}C_n$-$Pt_{SA}$ cocatalysts tend to have a strong affinity of hydrogen.

In contrast, with the introduction of surface decoration (-O, -F, and OH groups) on the prototype $Ti_{n+1}C_n$, the $Ti_{n+1}C_nT_x$-$Pt_{SA}$ has a strongly weakened adsorption of H which leads to a higher $\Delta G_{*H}$ (Figure 4(b-c)). For instance, the $\Delta G_{*H}$ for $Ti_4C_3O_2$-$Pt_{SA}$, $Ti_4C_3F_2$-$Pt_{SA}$ and $Ti_4C_3(OH)_2$-$Pt_{SA}$ are -0.53, +0.59 and +0.32 eV. Interestingly, different from the naked MXene $Ti_{n+1}C_n$-$Pt_{SA}$, there is a strong dependence of the value of $\Delta G_{*H}$ on the thickness for $Ti_{n+1}C_nT_x$-$Pt_{SA}$. For instance, when the thickness of these MXenes with O, F and OH terminations is reduced from $n$ = 3 to 1, the value of $\Delta G_{*H}$ increases and approaches the thermoneutral condition, i.e., especially for $Ti_2CH_2O_2$-$Pt_{SA}$ ($\Delta G_{*H}$ = +0.01 eV). These calculated results suggest a strong effect of quantum confinement on the adsorbing energetics of hydroge, likely due to the more tightly confined states and alter the electron affinity of MXene. The stronger uptake of hydrogen with reduced thickness of F- and OH- terminated MXene, while an insensitive change is found for O-terminated MXene. For O and F terminated surfaces, the HER process even becomes thermodynamically uphill, implying a reduced affinity to hydrogen. Our work suggests that the HER catalytic activity of the MXene is not solely governed by the local environment of the surface, the thickness of cationic framework plays an important role. Our result suggests that the surface activity of MXene based derivatives can be promoted by adjusting the thickness of MXene to take advantage of the quantum confinement effect.

### 3.3. Interfacial charge transfer and electronic properties

To identify the trend of the adsorption of H on $Ti_{n+1}C_nT_x$-$Pt_{SA}$ ($n$ = 1, 2 and 3, T = F and OH), the scaling relation between charge transfer of Pt atoms and $\Delta G_{*H}$ is plotted in Figure 4(e). The compounds of $Ti_{n+1}C_nO_2$-$Pt_{SA}$ are not considered in this diagram. As shown in Figure 3, H is more energetically favorable to be adsorbed on the O atoms near Pt on $Ti_{n+1}C_nO_2$-$Pt_{SA}$. For the others there is a nearly linear relationship between charge transfer and the $\Delta G_{*H}$ with a good correlation coefficient $R^2$ = 0.86. The affinity of H toward Pt increases with the amount of negative charge at Pt. The optimal charge of Pt for a thermoneutral adsorption is around -0.3 e and a larger charge transfer from the host sheet to Pt atoms leads to a stronger adsorption of H, deteriorating the HER performance.

The above correlation is generally found in other types of substitutionally Pt-doped MXenes with different terminations (see Table 2), which provides a clue into the analysis of HER activity via judging the charged states of catalytic site. In addition, we also calculate the work functions ($W$) of all $Ti_{n+1}C_nT_x$-$Pt_{SA}$ by $W = E_{vac}$ - $E_F$, where $E_{vac}$ is the energy of vacuum level and $E_F$ is the Fermi level. The work function is the minimum energy needed to remove an electron from the surface toward a vacuum level. Figure 5 shows the scatter plot between the work function and $\Delta G_{*H}$, and there is no direct correlation of these two variables with a correlation coefficient $R^2$ = 0.083. The results indicate that catalytic activity of HER is highly correlated with the local chemical environment of the active site, rather than the global electrostatic properties of the surface of monolayers.

Furthermore, we analyze the spatial distribution of charge transfer of the active site through calculating the differential charge density (Figure 6). The electrons are transferred from active site to the adsorbed H. It is known that the d-orbitals of the active site play a critical role in the catalytic performance [51]. The projected DOS as shown in Figure 7 indicates that the d orbitals of Pt are coupling with the s orbital of H, especially true for $Ti_2CH_2F_2$-$Pt_{SA}$. This scenario is absent in the $Ti_2CH_2O_2$-$Pt_{SA}$ because the catalytic site of it is O rather than Pt with the absence of the d states (Figure 7(a)). This could enhance the hydrogen-platinum interaction via the hybridization of d-orbitals of Pt and s-orbital of H.

### 3.4. Strain and solvent effects

Strain engineering is an effective approach to modulate the electronic properties for promoting the catalytic performance [19, 20, 38, 52]. Hence, we explored the effect of strain on the activity of three promising catalysts, which is shown in Figure 8(a-c). The calculated $\Delta G_{*H}$ increases from -0.05 eV at -2% compressive strain to 0.60 eV at 6% tensile strain on $Ti_2CF_2$-$Pt_{SA}$. This indicates that applying strain, though residual strain existing around the edge, can be a viable means for modulating the activity.

Solvent effect also plays a non-negligible role on chemical reactions at the water/solid interface [53, 54]. Water can affect the catalytic process as a competitive adsorbate. Therefore, the explicit solvent method is carried out to explore the effect for HER on three promising catalysts. As shown in Figure 8(d), the explicit solvent model is proved to play a minor role in HER for $Ti_2CF_2$-$Pt_{SA}$ and $Ti_2CH_2O_2$-$Pt_{SA}$ within 0.1 eV. But for $Ti_2CO_2$-$Pt_{SA}$, H tends to escape from the surface to form $H_3O^+$ with a $H_2O$ molecule, then form $H_5O_2^+$ with another $H_2O$ molecule by hydrogen bonding (Figure 8(d)). This suggests that the performance of $Ti_2CO_2$-$Pt_{SA}$ may be compromised in the presence of water molecule.

### 3.5. Multistep HER mechanism and thermal stability

Although experimental and theoretical studies have proved that $\Delta G_{*H}$ is the best descriptor for the HER, the actual process is more complex than this situation. The HER is a multistep electrochemical reaction [19, 52] as shown in Figure 9(a). The first step is the H adsorption on the surface of catalyst, namely Volmer reaction ($H^+ + e^- \rightarrow {}^*H$). Then the release of $H_2$ can be achieved through two mechanisms, namely Heyrovsky and Tafel reactions. For Heyrovsky mechanism, ${}^*H$ combines with $H^+$ and an electron to form a $H_2$ molecule (${}^*H + H^+ + e^- \rightarrow H_2$). Alternatively, $H_2$ could be formed by the combination of two ${}^*H$ on the surface of catalyst (Tafel step, $2{}^*H \rightarrow H_2$). To reveal the reaction barrier of HER, the climbed nudged elastic band (CI-NEB) [55] method is used to calculate the transition state of three catalysts. The initial state (IS), transition state (TS) and final state (FS) with the corresponding energy barriers of two mechanisms are shown in Figure 9(b-d). On the $Ti_2CF_2$-$Pt_{SA}$, it can be found that the energy barrier of Heyrovsky reaction is 2.05 eV, which is lower than that of the Tafel reaction (4.01 eV). Similarly, the energy barrier on $Ti_2CH_2O_2$-$Pt_{SA}$ by Heyrovsky mechanism is 0.76 eV lower than that of Tafel mechanism (2.19 eV). This result indicates that both $Ti_2CF_2$-$Pt_{SA}$ and $Ti_2CH_2O_2$-$Pt_{SA}$ tend to generate $H_2$ by Volmer-Heyrovsky mechanism.

Since $Ti_2CF_2$-$Pt_{SA}$ and $Ti_2CH_2O_2$-$Pt_{SA}$ are proved to be promising for HER catalysts, we next further examine the stability of the adsorbed Pt single atom which is critical for application. Besides, $Ti_2CO_2$-$Pt_{SA}$ is also examined for comparison. It is important that the Ti vacancies in these MXenes can dynamically stabilize Pt single atom at finite temperature. In our previous study [56], the thermal stability may deteriorate even though the binding of the atomic species is strong. Therefore, AIMD simulations should be carried out to validate the thermal stability of these co-catalysts despite a strong binding energies (Figure 2(c) and Table 2). The AIMD simulations are carried for the three systems at 350 K. At first, the velocities of ions are randomly initialized according to the Maxwell-Boltzmann distribution. Then, the systems are simulated under NVT ensemble at a constant temperature in the Nosé-Hoover thermostat [48]. The NVT process is simulated for 10 ps with a timestep of 1 fs. As shown in Figure 10, there is no obvious distortion for the three structures after AIMD simulations. The temperature and energy are oscillating near the equilibrium values, indicative of a high thermal stability of these systems. The comparison of HER activity for $Ti_2CF_2$-$Pt_{SA}$, $Ti_2CH_2O_2$-$Pt_{SA}$ and other reported MXene catalysts is illustrated in Table 4. As we can see, the $\Delta G_{*H}$ of $Ti_2CH_2O_2$-$Pt_{SA}$ is the smallest among these reported MXene-based HER catalysts, implying that the substitutional Pt-doped strategy is an effective approach to improve the HER activity of MXenes.

## 4. Conclusions

In this work, the catalytic activity of a series of $Ti_{n+1}C_nT_x$-$Pt_{SA}$ co-catalysts is systematically investigated by means of *ab initio* calculations. By taking three possible terminations (-O, -F and -OH) of MXenes into account, our results reveal that the terminations and thickness of MXenes can strongly affect the local HER activity of the anchoring Pt single atom. Interestingly, the HER performance is significantly promoted with the thickness of $Ti_{n+1}C_nT_x$ reducing from $n = 3$ to 1. Through adopting this strategy, $Ti_2CO_2$-$Pt_{SA}$, $Ti_2CF_2$-$Pt_{SA}$ and $Ti_2CH_2O_2$-$Pt_{SA}$ are proved to exhibit high efficient HER performance with the $\Delta G_{*H}$ of nearly zero. However, $Ti_2CF_2$ and $Ti_2CH_2O_2$ are easier to form Ti vacancies to anchor Pt atoms with much smaller formation energy of 3.60 and 3.49 eV than $Ti_2CO_2$ (9.33 eV), indicative of the possibility of experimental realization. Although $Ti_2CO_2$-$Pt_{SA}$, $Ti_2CF_2$-$Pt_{SA}$ and $Ti_2CH_2O_2$-$Pt_{SA}$ remain thermodynamically stable at 350 K, the H adsorbed on $Ti_2CO_2$-$Pt_{SA}$ will form $H_5O_2^+$ with $H_2O$ molecule, which hinders the HER process above $Ti_2CO_2$-$Pt_{SA}$. Our work suggests that the HER catalytic activity of the MXene derivatives is not solely governed by the local environment of the surface, and the thickness of cationic framework plays an important role as well. Our findings are useful for guiding experiments for enhancing the HER activity of Pt-doped MXenes via tailoring surface terminations and thickness engineering.

**Data availability statement**

All data that support the findings of this study are included within the article (and any supplementary files).

## Acknowledgments

This work is supported by the Natural Science Foundation of China (Grant 22022309) and Natural Science Foundation of Guangdong Province, China (2021A1515010024), the University of Macau (SRG2019-00179-IAPME, MYRG2020-00075-IAPME) and the Science and Technology Development Fund from Macau SAR (FDCT-0163/2019/A3). This work was performed in part at the High-Performance Computing Cluster (HPCC) which is supported by Information and Communication Technology Office (ICTO) of the University of Macau.

## Author contributions

Zheng Shu: Design, calculation and writing the original draft.

Yongqing Cai: Supervision, funding acquisition and editing the manuscript.

## Conflict of interest

There are no conflicts of interest to declare.

## ORCID iDs

Zheng Shu  https://orcid.org/0000-0002-7361-4084

Yongqing Cai  https://orcid.org/0000-0002-3565-574X

**Table 1.** Lattice constants ($a = b$), layer thickness ($l$) of primitive cell of $Ti_{n+1}C_nT_x$ and defect formation energies ($E_f$) of 3 × 3 × 1 supercell. The layer thickness is calculated

by atomic covalent radius.

|  | $a$ (Å) | $l$ (Å) | $E_f$ (eV) |
|---|---|---|---|
| $Ti_2C$ | 3.05 | 5.01 | 2.73 |
| $Ti_2CO_2$ | 3.01 | 5.72 | 9.33 |
| $Ti_2CF_2$ | 3.03 | 6.24 | 3.60 |
| $Ti_2CH_2O_2$ | 3.04 | 7.53 | 3.49 |
| $Ti_3C_2$ | 3.06 | 7.43 | 2.77 |
| $Ti_3C_2O_2$ | 3.01 | 8.23 | 7.44 |
| $Ti_3C_2F_2$ | 3.05 | 8.70 | 4.10 |
| $Ti_3C_2H_2O_2$ | 3.06 | 9.98 | 4.04 |
| $Ti_4C_3$ | 3.07 | 9.83 | 1.93 |
| $Ti_4C_3O_2$ | 3.02 | 10.69 | 9.12 |
| $Ti_4C_3F_2$ | 3.06 | 11.08 | 3.26 |
| $Ti_4C_3H_2O_2$ | 3.06 | 12.35 | 3.02 |

**Table 2.** Binding energies ($E_b$) and charge transfer ($\Delta q$) of Pt single atoms, and work function ($W$) of $Ti_{n+1}C_nT_x$-$Pt_{SA}$. It should be noted that a negative $\Delta q$ indicates electrons transfer from the host sheet to Pt single atoms.

| Host | $E_b$ (eV) | $\Delta q$ (e) | $W$ (eV) |
|---|---|---|---|
| $Ti_2C$-$Pt_{SA}$ | -8.08 | -0.80 | 4.18 |
| $Ti_2CO_2$-$Pt_{SA}$ | -9.31 | +0.80 | 6.00 |
| $Ti_2CF_2$-$Pt_{SA}$ | -4.92 | -0.05 | 5.15 |
| $Ti_2CH_2O_2$-$Pt_{SA}$ | -5.50 | -0.30 | 2.11 |
| $Ti_3C_2$-$Pt_{SA}$ | -8.22 | -0.67 | 4.43 |
| Ti$_3$C$_2$O$_2$-Pt$_{SA}$ | -7.67 | +0.94 | 6.16 |
| Ti$_3$C$_2$F$_2$-Pt$_{SA}$ | -5.21 | +0.01 | 5.21 |
| Ti$_3$C$_2$H$_2$O$_2$-Pt$_{SA}$ | -5.55 | -0.21 | 1.96 |
| Ti$_4$C$_3$-Pt$_{SA}$ | -8.25 | -0.66 | 4.13 |
| Ti$_4$C$_3$O$_2$-Pt$_{SA}$ | -9.07 | +0.94 | 6.05 |
| Ti$_4$C$_3$F$_2$-Pt$_{SA}$ | -5.66 | 0.00 | 5.13 |
| Ti$_4$C$_3$H$_2$O$_2$-Pt$_{SA}$ | -5.84 | -0.24 | 2.08 |

**Table 3.** The calculated vibrational frequencies, zero-point energies ($E_{ZPE}$), entropy corrections ($TS$) for adsorbed H on Ti$_{n+1}$C$_n$T$_x$-Pt$_{SA}$, and the corresponding Gibbs free energy ($\Delta G_{*H}$).

| Host | Vibrational Frequencies (cm$^{-1}$) | | | $E_{ZPE}$ (eV) | $TS$ (eV) | $\Delta G_{*H}$ (eV) |
|---|---|---|---|---|---|---|
| Ti$_2$C-Pt$_{SA}$ | 2093.53 | 206.11 | 184.81 | 0.15 | 0.06 | -0.57 |
| Ti$_2$CO$_2$-Pt$_{SA}$ | 3647.35 | 578.57 | 554.82 | 0.30 | 0.01 | -0.06 |
| Ti$_2$CF$_2$-Pt$_{SA}$ | 2140.41 | 544.96 | 530.52 | 0.20 | 0.01 | +0.04 |
| Ti$_2$CH$_2$O$_2$-Pt$_{SA}$ | 1574.10 | 874.05 | 867.93 | 0.21 | 0.004 | +0.01 |
| Ti$_3$C$_2$-Pt$_{SA}$ | 2134.50 | 187.25 | 179.21 | 0.16 | 0.06 | -0.64 |
| Ti$_3$C$_2$O$_2$-Pt$_{SA}$ | 3572.56 | 767.17 | 544.15 | 0.30 | 0.01 | -0.52 |
| Ti$_3$C$_2$F$_2$-Pt$_{SA}$ | 2157.30 | 518.07 | 511.85 | 0.20 | 0.02 | +0.32 |
| Ti$_3$C$_2$H$_2$O$_2$-Pt$_{SA}$ | 1576.20 | 866.41 | 860.38 | 0.20 | 0.004 | +0.20 |
| Ti$_4$C$_3$-Pt$_{SA}$ | 2115.87 | 147.54 | 121.72 | 0.13 | 0.00 | -0.61 |
| Ti$_4$C$_3$O$_2$-Pt$_{SA}$ | 3609.12 | 700.38 | 562.66 | 0.30 | 0.01 | -0.53 |
| Ti$_4$C$_3$F$_2$-Pt$_{SA}$ | 2240.02 | 416.01 | 402.80 | 0.19 | 0.02 | +0.59 |

| $Ti_4C_3H_2O_2$-$Pt_{SA}$ | 1429.79 | 861.28 | 854.86 | 0.20 | 0.004 | +0.32 |
|---|---|---|---|---|---|---|

**Table 4.** The comparison of $\Delta G_{*H}$ of our work with other reported MXenes.

| Catalysts | $\Delta G_{*H}$ (eV) | Catalysts | $\Delta G_{*H}$ (eV) |
|---|---|---|---|
| $Ti_2CF_2$-$Pt_{SA}$ (this work) | +0.04 | $Ni$-$Cr_2CO_2$ | +0.06 |
| $Ti_2CH_2O_2$-$Pt_{SA}$ (this work) | +0.01 | $Co$-$Cr_2CO_2$ | +0.07 |
| $Ti_2CT_x$ | +0.358 | $Au$-$Cr_2TiC_2O_{2-\delta}$ | +0.07 |
| $Mo_2CT_x$ | +0.048 | $Co$-$Cr_2TiC_2O_{2-\delta}$ | -0.07 |
| $Mo_2TiC_2T_x$-$Pt_{SA}$ | -0.08 | $Ni$-$Mo_2TiC_2O_{2-\delta}$ | +0.02 |
| $Mo_2TiC_2T_x$ | -0.19 | $Ti_3CNO_2$ | -0.165 |
| $Pt_O$-$Mo_2TiC_2O_2$ | +0.02 | $V_3CNO_2$ | +0.077 |

References: [32, 33, 35-37, 39-41].

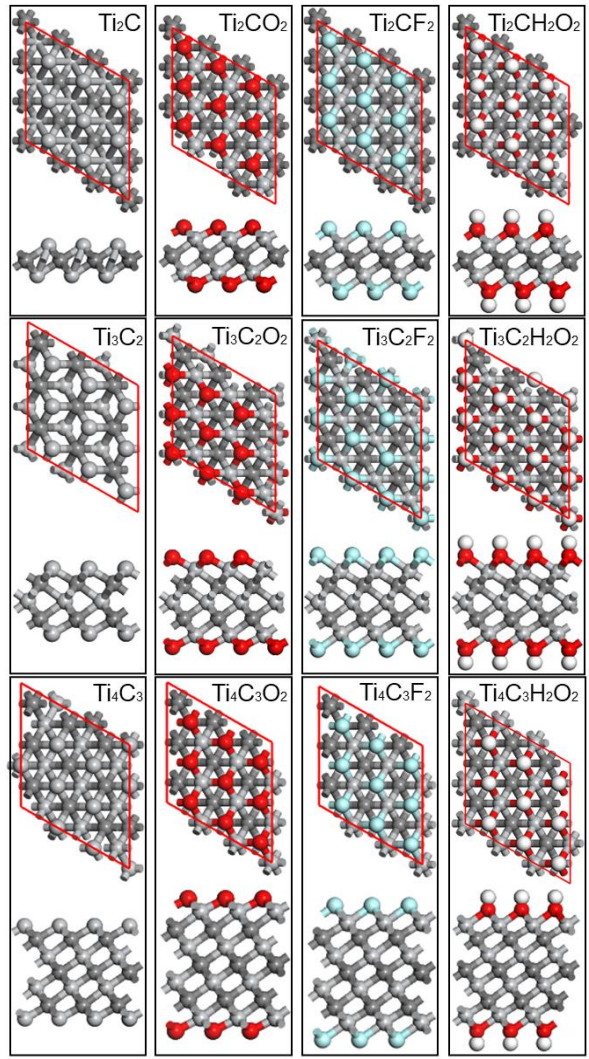

**Figure 1.** Configuration view of 3 × 3 × 1 supercell of Ti$_{n+1}$C$_n$T$_x$ ($n$ = 1, 2 and 3, T$_x$ = O, F and OH). The light grey, dark grep, red, cyan and white balls represent Ti, C, O, F and H atoms, respectively.

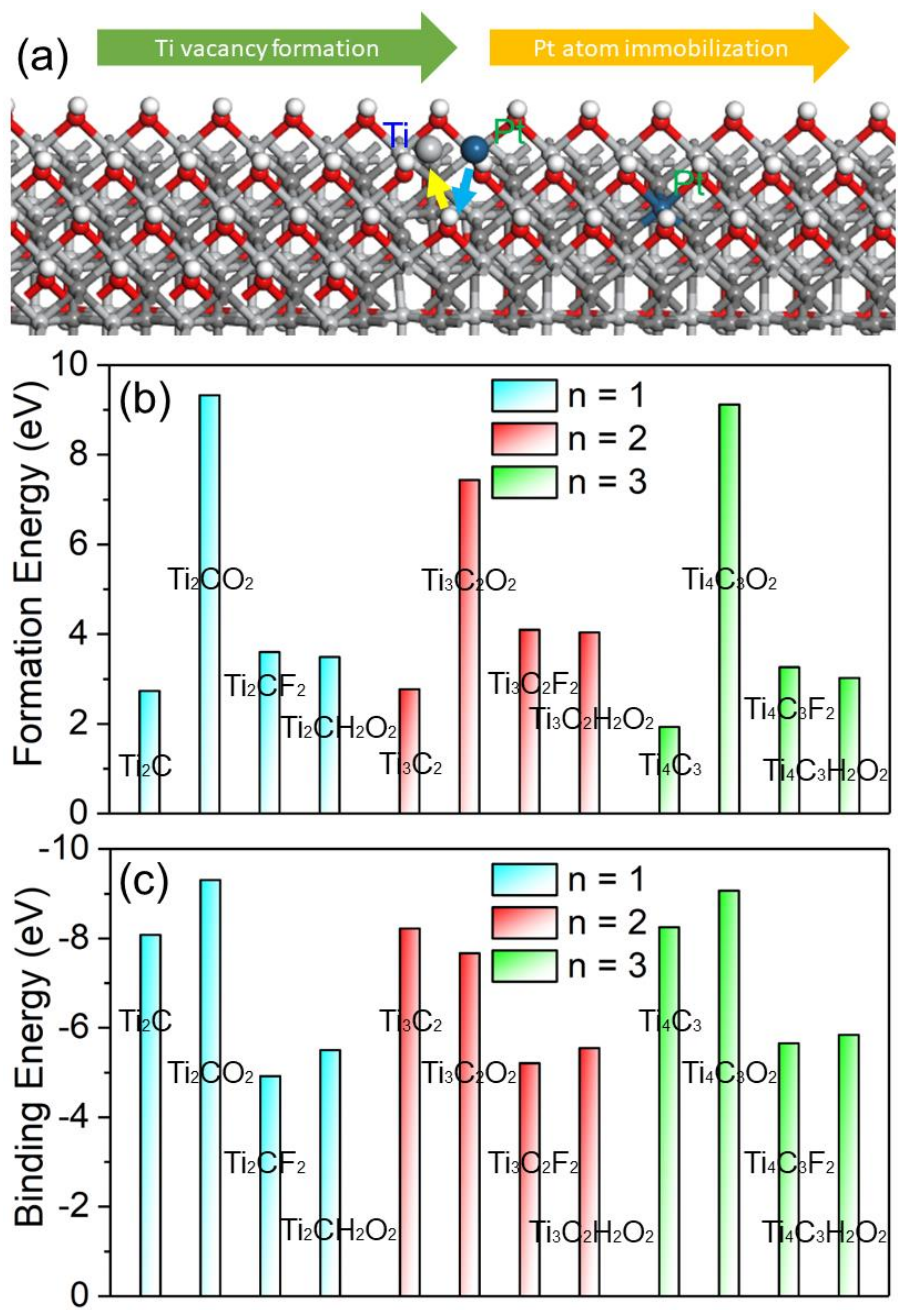

**Figure 2.** (a) The flow diagram of the synthesis mechanism for $Ti_{n+1}C_nT_x$-$Pt_{SA}$ during the HER process. Histogram of calculated (b) defect formation energies of Ti vacancy and (c) binding energies of Pt above Ti vacancy site.

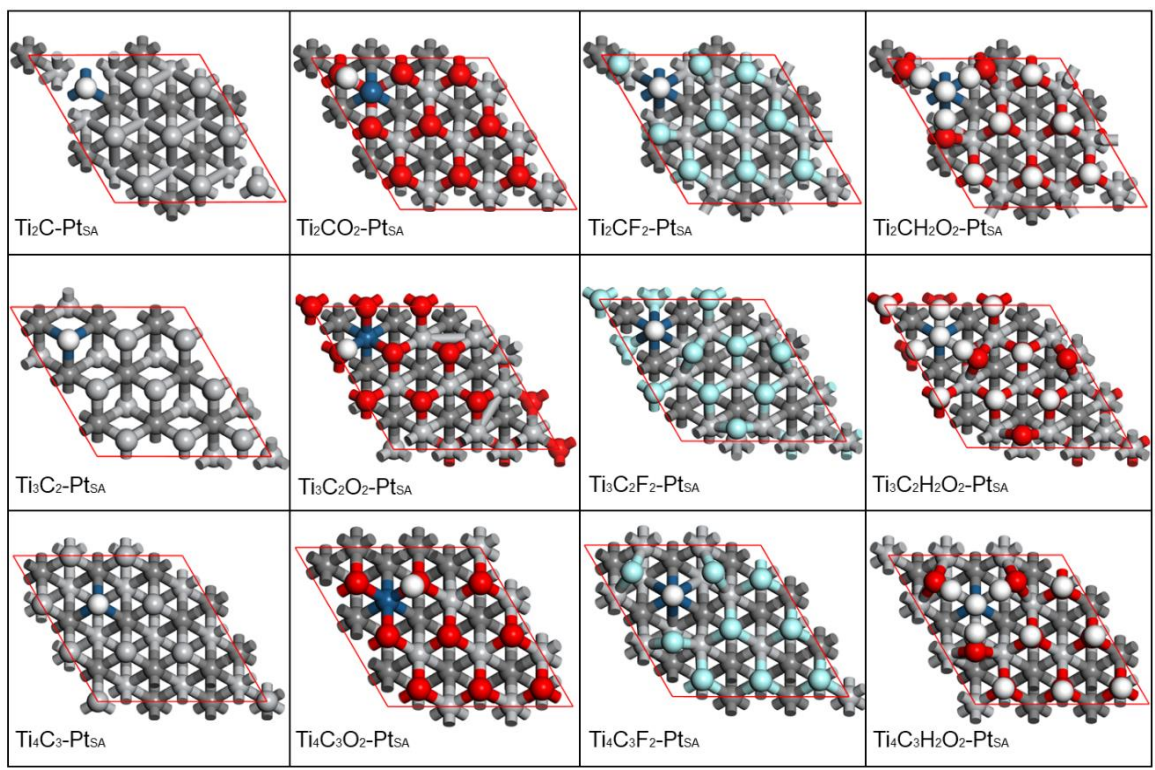

**Figure 3.** The atomic structures of $Ti_{n+1}C_nT_x$-$Pt_{SA}$ ($n$ = 1, 2 and 3, $T_x$ = O, F and OH) with the adsorbed $H^+$. The light grey, dark grep, red, cyan, white and navy balls represent Ti, C, O, F, H and Pt atoms, respectively.

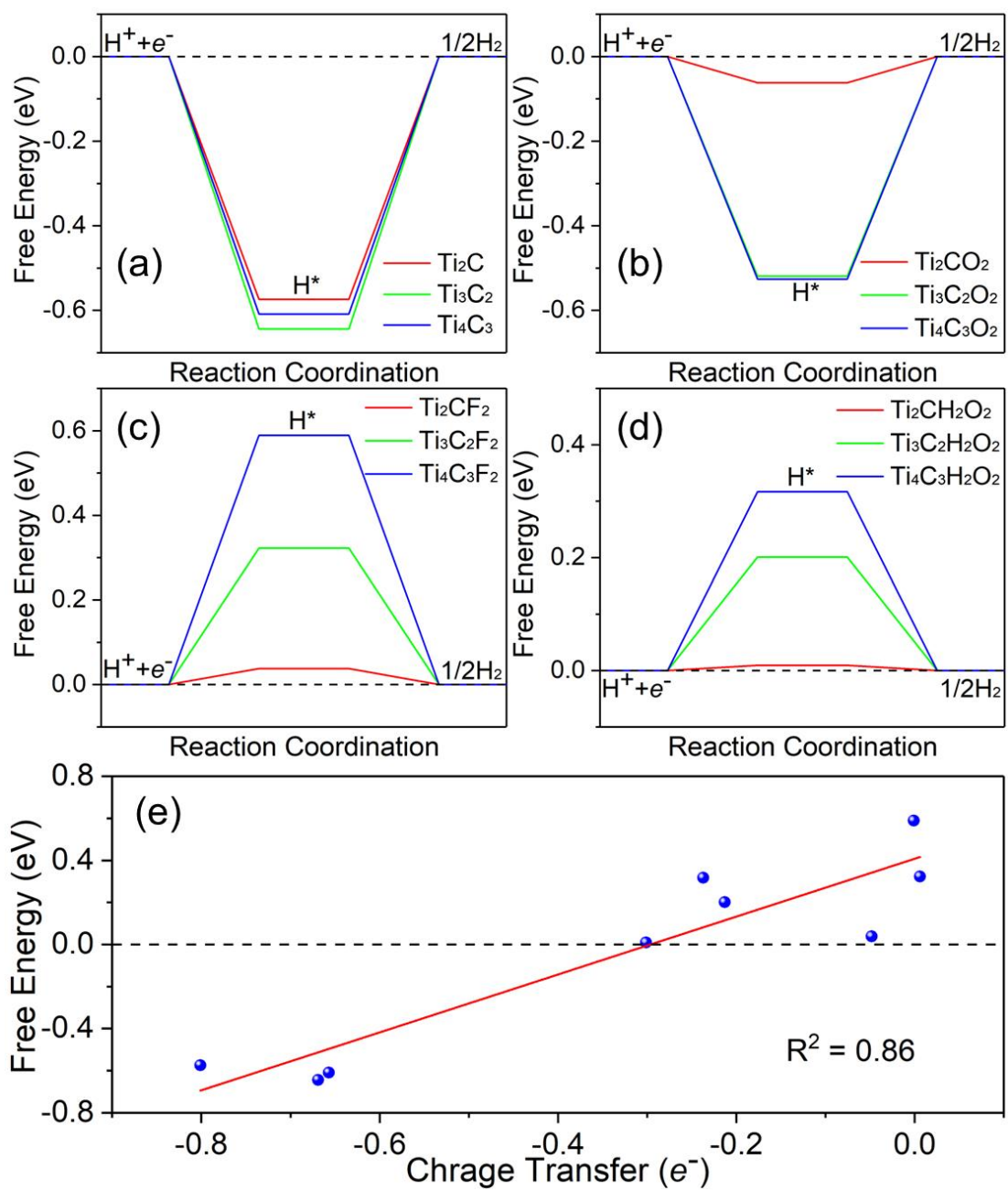

**Figure 4.** Gibbs free energy changes for hydrogen reduction reaction on (a) $Ti_{n+1}C_n$-$Pt_{SA}$, (b) $Ti_{n+1}C_nO_2$-$Pt_{SA}$, (c) $Ti_{n+1}C_nF_2$-$Pt_{SA}$ and (d) $Ti_{n+1}C_nH_2O_2$-$Pt_{SA}$. (e) Linear scaling fitting between the amount of the charge transfer (charge state) of Pt atom and the $\Delta G_{*H}$. Noted that $Ti_{n+1}C_nO_2$-$Pt_{SA}$ is ruled out because H adsorbed on O rather than Pt for these compounds.

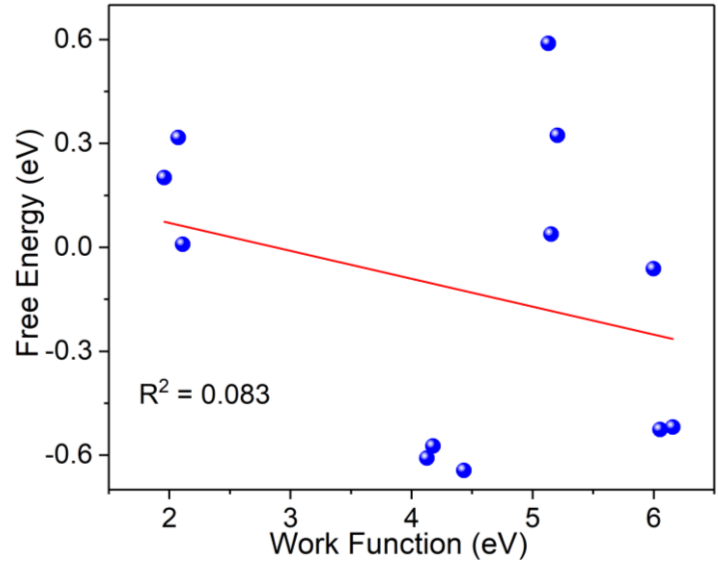

**Figure 5.** The poor scaling relationship between work function of $Ti_{n+1}C_nT_x$-$Pt_{SA}$ surface and $\Delta G_{*H}$.

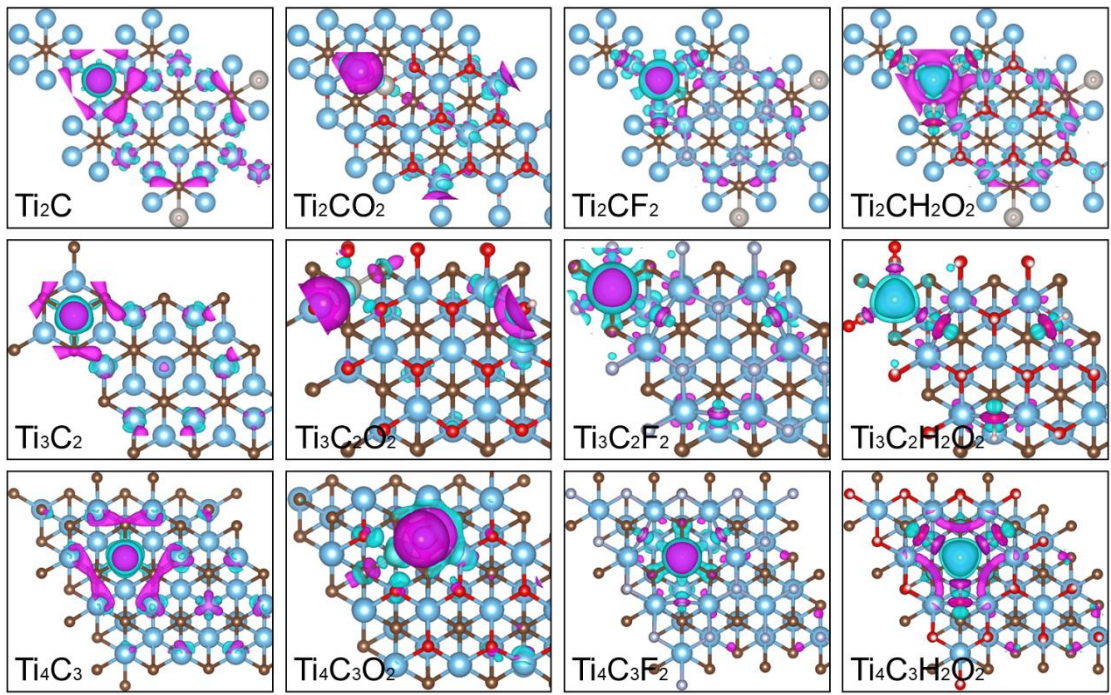

**Figure 6.** Iso-surface plot of differential charge density of $Ti_{n+1}C_nT_x$ ($n$ = 1, 2 and 3, T = O, F and OH). The red (blue) color represents the accumulation (depletion) of electrons. The iso-surface level is set to 0.0015 e Å⁻³.

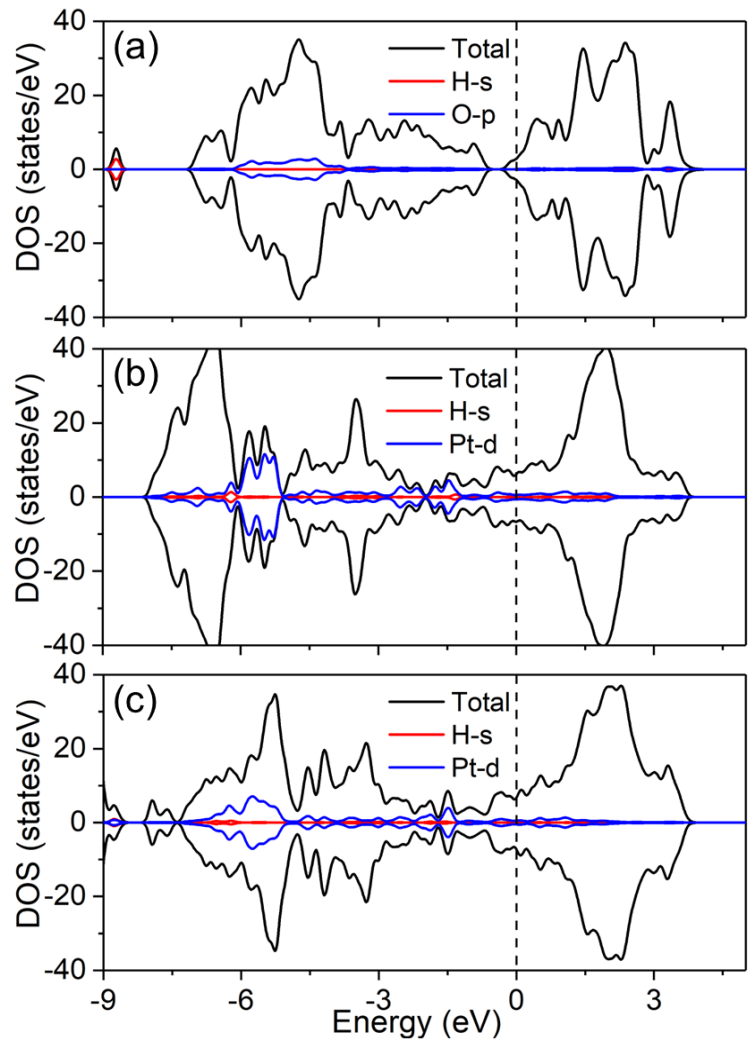

**Figure 7.** Total DOS and projected DOS of s-orbital of H and d-orbitals of Pt for H

adsorbed (a) $Ti_2CO_2$-$Pt_{SA}$, (b) $Ti_2CF_2$-$Pt_{SA}$ and (c) $Ti_2CH_2O_2$-$Pt_{SA}$. The Fermi level is set to zero.

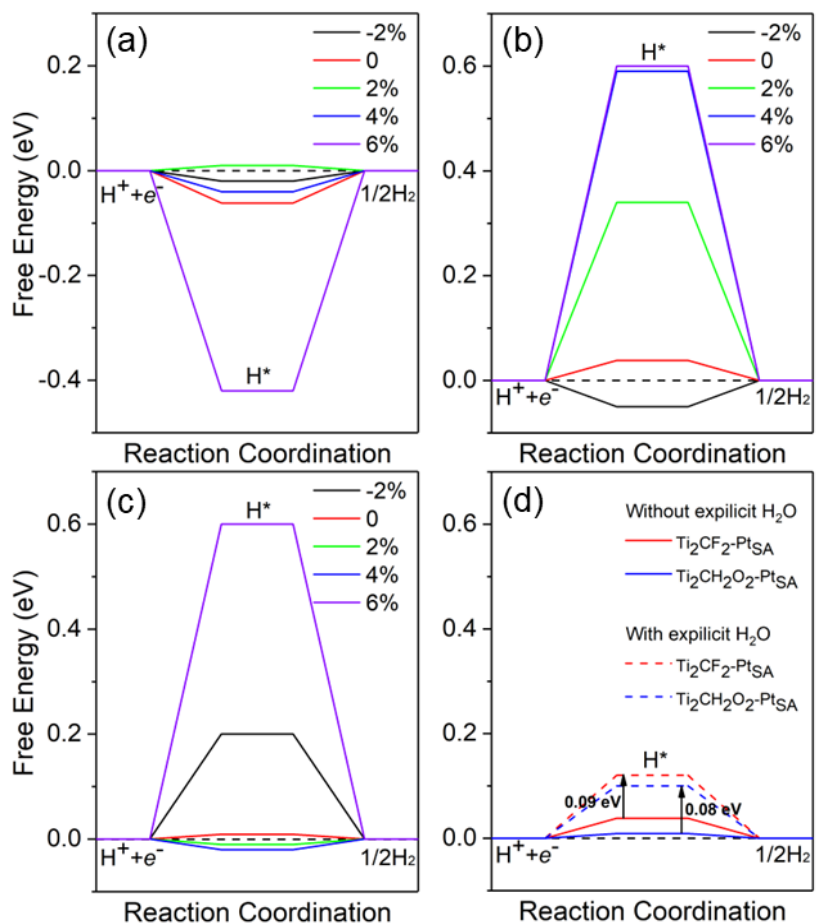

**Figure 8.** Gibbs free energy changes for hydrogen reduction reaction on (a) $Ti_2CO_2$-$Pt_{SA}$, (b) $Ti_2CF_2$-$Pt_{SA}$ and (c) $Ti_2CH_2O_2$-$Pt_{SA}$ under biaxial strain from -2% to 6%. (d) The comparison of Gibbs free energy changes for $Ti_2CF_2$-$Pt_{SA}$ and $Ti_2CH_2O_2$-$Pt_{SA}$ in the presence of explicit water.

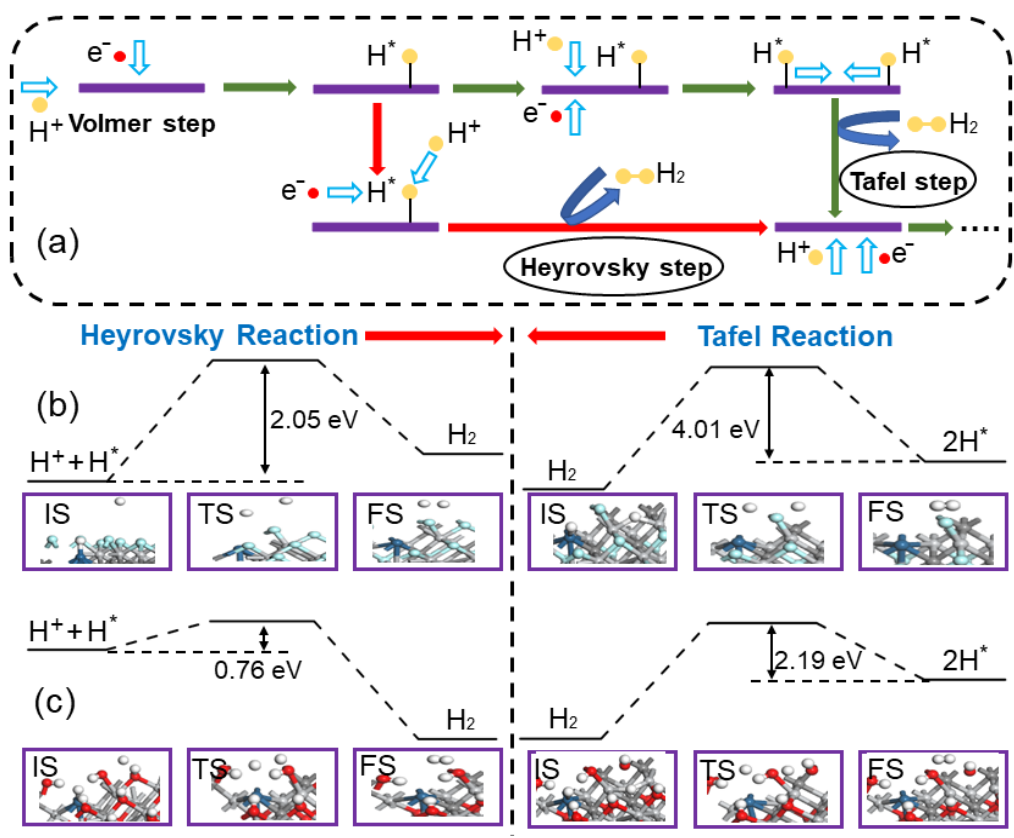

**Figure 9.** (a) Schematic pathways for the hydrogen evolution reaction. Energy profiles for the Heyrovsky and Tafel mechanisms for (b) $Ti_2CF_2$-$Pt_{SA}$ and (c) $Ti_2CH_2O_2$-$Pt_{SA}$.

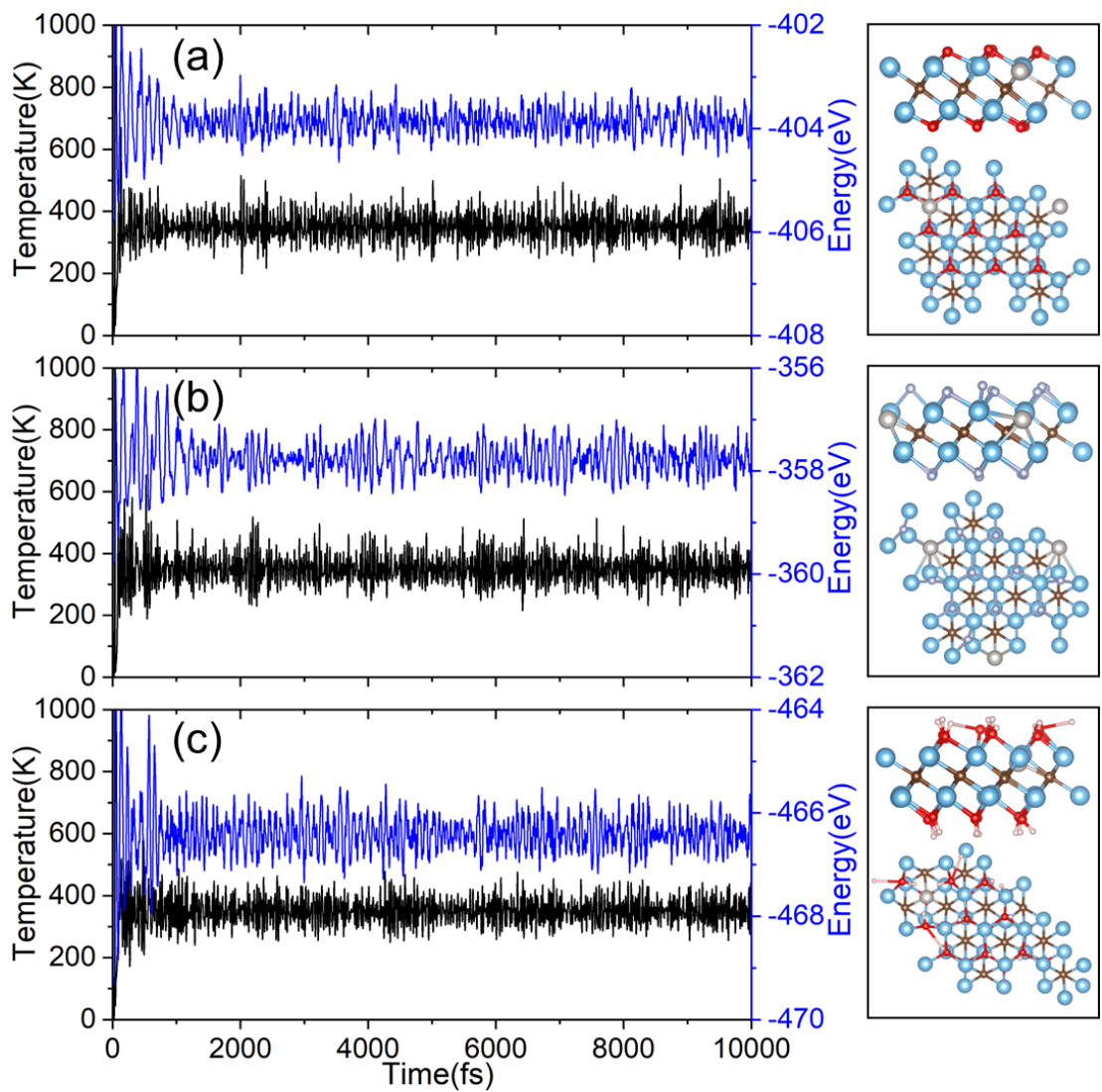

**Figure 10.** AIMD simulations of (a) $Ti_2CO_2$-$Pt_{SA}$, (b) $Ti_2CF_2$-$Pt_{SA}$ and (c) $Ti_2CH_2O_2$-$Pt_{SA}$ at temperature of 350 K.